\documentclass[longauth]{aa}  
\usepackage{graphicx}
\usepackage{txfonts}
\usepackage{pdflscape}
\usepackage{soul}
\usepackage{adjustbox}
\usepackage{array}
\usepackage{lscape}
\usepackage{bm}
\usepackage{natbib}
\bibpunct{(}{)}{;}{a}{}{,} 
\usepackage{euclid}

\usepackage[colorlinks=true,urlcolor=blue,citecolor=blue,linkcolor=blue]{hyperref}

\begin{document}

   \title{UNIONS and \textit{Euclid}: Calibrating point source--galaxy separation for ground-based surveys\thanks{This paper is published on behalf of the Euclid Consortium.}}

\newcommand{\orcid}[1]{}   
\author{G.~Thomas\orcid{0000-0002-2468-5521}\thanks{\email{gthomas@iac.es}}\inst{\ref{aff1},\ref{aff2}}
\and S.~E.~T.~Smith\orcid{0000-0002-6946-8280}\inst{\ref{aff3}}
\and S.~Gwyn\orcid{0000-0001-8221-8406}\inst{\ref{aff4}}
\and A.~W.~McConnachie\orcid{0000-0003-4666-6564}\inst{\ref{aff4}}
\and J.-C.~Cuillandre\orcid{0000-0002-3263-8645}\inst{\ref{aff5}}
\and M.~Gatto\orcid{0000-0003-4636-6457}\inst{\ref{aff6}}
\and E.~Balbinot\orcid{0000-0002-1322-3153}\inst{\ref{aff7},\ref{aff8}}
\and G.~Battaglia\orcid{0000-0002-6551-4294}\inst{\ref{aff2},\ref{aff1}}
\and T.~de~Boer\orcid{0000-0001-5486-2747}\inst{\ref{aff9}}
\and V.~Dornan\orcid{0000-0002-7731-1291}\inst{\ref{aff10}}
\and C.~M.~Gutierrez\orcid{0000-0001-7854-783X}\inst{\ref{aff2},\ref{aff1}}
\and A.~Lan\c{c}on\orcid{0000-0002-7214-8296}\inst{\ref{aff11}}
\and N.~F.~Martin\orcid{0000-0002-1349-202X}\inst{\ref{aff11},\ref{aff12}}
\and G.~E.~Medina\orcid{0000-0003-0105-9576}\inst{\ref{aff13}}
\and J.~F.~Navarro\orcid{0000-0003-3862-5076}\inst{\ref{aff3}}
\and D.~Scott\orcid{0000-0002-6878-9840}\inst{\ref{aff14}}
\and M.~Urbano\orcid{0000-0001-5640-0650}\inst{\ref{aff15},\ref{aff11}}
\and G.~Zamorani\orcid{0000-0002-2318-301X}\inst{\ref{aff16}}
\and B.~Altieri\orcid{0000-0003-3936-0284}\inst{\ref{aff17}}
\and S.~Andreon\orcid{0000-0002-2041-8784}\inst{\ref{aff18}}
\and N.~Auricchio\orcid{0000-0003-4444-8651}\inst{\ref{aff16}}
\and C.~Baccigalupi\orcid{0000-0002-8211-1630}\inst{\ref{aff19},\ref{aff20},\ref{aff21},\ref{aff22}}
\and M.~Baldi\orcid{0000-0003-4145-1943}\inst{\ref{aff23},\ref{aff16},\ref{aff24}}
\and S.~Bardelli\orcid{0000-0002-8900-0298}\inst{\ref{aff16}}
\and P.~Battaglia\orcid{0000-0002-7337-5909}\inst{\ref{aff16}}
\and A.~Biviano\orcid{0000-0002-0857-0732}\inst{\ref{aff20},\ref{aff19}}
\and M.~Bolzonella\orcid{0000-0003-3278-4607}\inst{\ref{aff16}}
\and E.~Branchini\orcid{0000-0002-0808-6908}\inst{\ref{aff25},\ref{aff26},\ref{aff18}}
\and M.~Brescia\orcid{0000-0001-9506-5680}\inst{\ref{aff27},\ref{aff6}}
\and S.~Camera\orcid{0000-0003-3399-3574}\inst{\ref{aff28},\ref{aff29},\ref{aff30}}
\and V.~Capobianco\orcid{0000-0002-3309-7692}\inst{\ref{aff30}}
\and C.~Carbone\orcid{0000-0003-0125-3563}\inst{\ref{aff31}}
\and V.~F.~Cardone\inst{\ref{aff32},\ref{aff33}}
\and J.~Carretero\orcid{0000-0002-3130-0204}\inst{\ref{aff34},\ref{aff35}}
\and M.~Castellano\orcid{0000-0001-9875-8263}\inst{\ref{aff32}}
\and G.~Castignani\orcid{0000-0001-6831-0687}\inst{\ref{aff16}}
\and S.~Cavuoti\orcid{0000-0002-3787-4196}\inst{\ref{aff6},\ref{aff36}}
\and K.~C.~Chambers\orcid{0000-0001-6965-7789}\inst{\ref{aff9}}
\and A.~Cimatti\inst{\ref{aff37}}
\and C.~Colodro-Conde\inst{\ref{aff2}}
\and G.~Congedo\orcid{0000-0003-2508-0046}\inst{\ref{aff10}}
\and C.~J.~Conselice\orcid{0000-0003-1949-7638}\inst{\ref{aff38}}
\and L.~Conversi\orcid{0000-0002-6710-8476}\inst{\ref{aff39},\ref{aff17}}
\and Y.~Copin\orcid{0000-0002-5317-7518}\inst{\ref{aff40}}
\and F.~Courbin\orcid{0000-0003-0758-6510}\inst{\ref{aff41},\ref{aff42},\ref{aff43}}
\and H.~M.~Courtois\orcid{0000-0003-0509-1776}\inst{\ref{aff44}}
\and M.~Cropper\orcid{0000-0003-4571-9468}\inst{\ref{aff45}}
\and H.~Degaudenzi\orcid{0000-0002-5887-6799}\inst{\ref{aff46}}
\and G.~De~Lucia\orcid{0000-0002-6220-9104}\inst{\ref{aff20}}
\and H.~Dole\orcid{0000-0002-9767-3839}\inst{\ref{aff47}}
\and F.~Dubath\orcid{0000-0002-6533-2810}\inst{\ref{aff46}}
\and X.~Dupac\inst{\ref{aff17}}
\and M.~Fabricius\orcid{0000-0002-7025-6058}\inst{\ref{aff48},\ref{aff49}}
\and M.~Farina\orcid{0000-0002-3089-7846}\inst{\ref{aff50}}
\and R.~Farinelli\inst{\ref{aff16}}
\and S.~Farrens\orcid{0000-0002-9594-9387}\inst{\ref{aff5}}
\and S.~Ferriol\inst{\ref{aff40}}
\and F.~Finelli\orcid{0000-0002-6694-3269}\inst{\ref{aff16},\ref{aff51}}
\and S.~Fotopoulou\orcid{0000-0002-9686-254X}\inst{\ref{aff52}}
\and N.~Fourmanoit\orcid{0009-0005-6816-6925}\inst{\ref{aff53}}
\and M.~Frailis\orcid{0000-0002-7400-2135}\inst{\ref{aff20}}
\and E.~Franceschi\orcid{0000-0002-0585-6591}\inst{\ref{aff16}}
\and M.~Fumana\orcid{0000-0001-6787-5950}\inst{\ref{aff31}}
\and S.~Galeotta\orcid{0000-0002-3748-5115}\inst{\ref{aff20}}
\and K.~George\orcid{0000-0002-1734-8455}\inst{\ref{aff54}}
\and B.~Gillis\orcid{0000-0002-4478-1270}\inst{\ref{aff10}}
\and C.~Giocoli\orcid{0000-0002-9590-7961}\inst{\ref{aff16},\ref{aff24}}
\and J.~Gracia-Carpio\orcid{0000-0003-4689-3134}\inst{\ref{aff48}}
\and A.~Grazian\orcid{0000-0002-5688-0663}\inst{\ref{aff55}}
\and F.~Grupp\inst{\ref{aff48},\ref{aff49}}
\and S.~V.~H.~Haugan\orcid{0000-0001-9648-7260}\inst{\ref{aff56}}
\and W.~Holmes\orcid{0009-0007-8554-4646}\inst{\ref{aff57}}
\and I.~M.~Hook\orcid{0000-0002-2960-978X}\inst{\ref{aff58}}
\and F.~Hormuth\inst{\ref{aff59}}
\and A.~Hornstrup\orcid{0000-0002-3363-0936}\inst{\ref{aff60},\ref{aff61}}
\and K.~Jahnke\orcid{0000-0003-3804-2137}\inst{\ref{aff12}}
\and M.~Jhabvala\inst{\ref{aff62}}
\and B.~Joachimi\orcid{0000-0001-7494-1303}\inst{\ref{aff63}}
\and S.~Kermiche\orcid{0000-0002-0302-5735}\inst{\ref{aff53}}
\and A.~Kiessling\orcid{0000-0002-2590-1273}\inst{\ref{aff57}}
\and M.~Kilbinger\orcid{0000-0001-9513-7138}\inst{\ref{aff5}}
\and B.~Kubik\orcid{0009-0006-5823-4880}\inst{\ref{aff40}}
\and M.~K\"ummel\orcid{0000-0003-2791-2117}\inst{\ref{aff49}}
\and M.~Kunz\orcid{0000-0002-3052-7394}\inst{\ref{aff64}}
\and H.~Kurki-Suonio\orcid{0000-0002-4618-3063}\inst{\ref{aff65},\ref{aff66}}
\and A.~M.~C.~Le~Brun\orcid{0000-0002-0936-4594}\inst{\ref{aff67}}
\and S.~Ligori\orcid{0000-0003-4172-4606}\inst{\ref{aff30}}
\and P.~B.~Lilje\orcid{0000-0003-4324-7794}\inst{\ref{aff56}}
\and V.~Lindholm\orcid{0000-0003-2317-5471}\inst{\ref{aff65},\ref{aff66}}
\and I.~Lloro\orcid{0000-0001-5966-1434}\inst{\ref{aff68}}
\and G.~Mainetti\orcid{0000-0003-2384-2377}\inst{\ref{aff69}}
\and O.~Mansutti\orcid{0000-0001-5758-4658}\inst{\ref{aff20}}
\and O.~Marggraf\orcid{0000-0001-7242-3852}\inst{\ref{aff70}}
\and M.~Martinelli\orcid{0000-0002-6943-7732}\inst{\ref{aff32},\ref{aff33}}
\and N.~Martinet\orcid{0000-0003-2786-7790}\inst{\ref{aff71}}
\and F.~Marulli\orcid{0000-0002-8850-0303}\inst{\ref{aff72},\ref{aff16},\ref{aff24}}
\and R.~J.~Massey\orcid{0000-0002-6085-3780}\inst{\ref{aff73}}
\and E.~Medinaceli\orcid{0000-0002-4040-7783}\inst{\ref{aff16}}
\and S.~Mei\orcid{0000-0002-2849-559X}\inst{\ref{aff74},\ref{aff75}}
\and M.~Meneghetti\orcid{0000-0003-1225-7084}\inst{\ref{aff16},\ref{aff24}}
\and E.~Merlin\orcid{0000-0001-6870-8900}\inst{\ref{aff55}}
\and G.~Meylan\orcid{0000-0001-6503-0209}\inst{\ref{aff76}}
\and A.~Mora\orcid{0000-0002-1922-8529}\inst{\ref{aff77}}
\and M.~Moresco\orcid{0000-0002-7616-7136}\inst{\ref{aff72},\ref{aff16}}
\and L.~Moscardini\orcid{0000-0002-3473-6716}\inst{\ref{aff72},\ref{aff16},\ref{aff24}}
\and R.~Nakajima\orcid{0009-0009-1213-7040}\inst{\ref{aff70}}
\and C.~Neissner\orcid{0000-0001-8524-4968}\inst{\ref{aff78},\ref{aff35}}
\and S.-M.~Niemi\orcid{0009-0005-0247-0086}\inst{\ref{aff79}}
\and C.~Padilla\orcid{0000-0001-7951-0166}\inst{\ref{aff78}}
\and S.~Paltani\orcid{0000-0002-8108-9179}\inst{\ref{aff46}}
\and F.~Pasian\orcid{0000-0002-4869-3227}\inst{\ref{aff20}}
\and V.~Pettorino\orcid{0000-0002-4203-9320}\inst{\ref{aff79}}
\and A.~Pezzotta\orcid{0000-0003-0726-2268}\inst{\ref{aff18}}
\and S.~Pires\orcid{0000-0002-0249-2104}\inst{\ref{aff5}}
\and G.~Polenta\orcid{0000-0003-4067-9196}\inst{\ref{aff80}}
\and M.~Poncet\inst{\ref{aff81}}
\and L.~A.~Popa\inst{\ref{aff82}}
\and F.~Raison\orcid{0000-0002-7819-6918}\inst{\ref{aff48}}
\and A.~Renzi\orcid{0000-0001-9856-1970}\inst{\ref{aff83},\ref{aff84},\ref{aff16}}
\and J.~Rhodes\orcid{0000-0002-4485-8549}\inst{\ref{aff57}}
\and G.~Riccio\inst{\ref{aff6}}
\and F.~Rizzo\orcid{0000-0002-9407-585X}\inst{\ref{aff20}}
\and E.~Romelli\orcid{0000-0003-3069-9222}\inst{\ref{aff20}}
\and M.~Roncarelli\orcid{0000-0001-9587-7822}\inst{\ref{aff16}}
\and C.~Rosset\orcid{0000-0003-0286-2192}\inst{\ref{aff74}}
\and B.~Rusholme\orcid{0000-0001-7648-4142}\inst{\ref{aff85}}
\and R.~Saglia\orcid{0000-0003-0378-7032}\inst{\ref{aff49},\ref{aff48}}
\and Z.~Sakr\orcid{0000-0002-4823-3757}\inst{\ref{aff86},\ref{aff87},\ref{aff88}}
\and A.~G.~S\'anchez\orcid{0000-0003-1198-831X}\inst{\ref{aff48}}
\and D.~Sapone\orcid{0000-0001-7089-4503}\inst{\ref{aff89}}
\and B.~Sartoris\orcid{0000-0003-1337-5269}\inst{\ref{aff49},\ref{aff20}}
\and P.~Schneider\orcid{0000-0001-8561-2679}\inst{\ref{aff70}}
\and T.~Schrabback\orcid{0000-0002-6987-7834}\inst{\ref{aff90}}
\and A.~Secroun\orcid{0000-0003-0505-3710}\inst{\ref{aff53}}
\and E.~Sihvola\orcid{0000-0003-1804-7715}\inst{\ref{aff91}}
\and P.~Simon\inst{\ref{aff70}}
\and C.~Sirignano\orcid{0000-0002-0995-7146}\inst{\ref{aff83},\ref{aff84}}
\and G.~Sirri\orcid{0000-0003-2626-2853}\inst{\ref{aff24}}
\and L.~Stanco\orcid{0000-0002-9706-5104}\inst{\ref{aff84}}
\and P.~Tallada-Cresp\'{i}\orcid{0000-0002-1336-8328}\inst{\ref{aff34},\ref{aff35}}
\and A.~N.~Taylor\inst{\ref{aff10}}
\and H.~I.~Teplitz\orcid{0000-0002-7064-5424}\inst{\ref{aff92}}
\and I.~Tereno\orcid{0000-0002-4537-6218}\inst{\ref{aff93},\ref{aff94}}
\and S.~Toft\orcid{0000-0003-3631-7176}\inst{\ref{aff95},\ref{aff96}}
\and R.~Toledo-Moreo\orcid{0000-0002-2997-4859}\inst{\ref{aff97},\ref{aff98}}
\and F.~Torradeflot\orcid{0000-0003-1160-1517}\inst{\ref{aff35},\ref{aff34}}
\and I.~Tutusaus\orcid{0000-0002-3199-0399}\inst{\ref{aff99},\ref{aff100},\ref{aff87}}
\and J.~Valiviita\orcid{0000-0001-6225-3693}\inst{\ref{aff65},\ref{aff66}}
\and T.~Vassallo\orcid{0000-0001-6512-6358}\inst{\ref{aff20},\ref{aff54}}
\and Y.~Wang\orcid{0000-0002-4749-2984}\inst{\ref{aff85}}
\and J.~Weller\orcid{0000-0002-8282-2010}\inst{\ref{aff49},\ref{aff48}}
\and E.~Zucca\orcid{0000-0002-5845-8132}\inst{\ref{aff16}}
\and A.~Montoro\orcid{0000-0003-4730-8590}\inst{\ref{aff99},\ref{aff100}}
\and M.~Sereno\orcid{0000-0003-0302-0325}\inst{\ref{aff16},\ref{aff24}}}
                                                                                   
\institute{Universidad de La Laguna, Dpto. Astrof\'\i sica, E-38206 La Laguna, Tenerife, Spain\label{aff1}
\and
Instituto de Astrof\'{\i}sica de Canarias, E-38205 La Laguna, Tenerife, Spain\label{aff2}
\and
Department of Physics and Astronomy, University of Victoria, Victoria, BC V8P 5C2, Canada\label{aff3}
\and
National Research Council, Herzberg Astronomy and Astrophysics Research Centre, 5071 W. Saanich Rd. Victoria, BC, V9E 2E7, Canada\label{aff4}
\and
Universit\'e Paris-Saclay, Universit\'e Paris Cit\'e, CEA, CNRS, AIM, 91191, Gif-sur-Yvette, France\label{aff5}
\and
INAF-Osservatorio Astronomico di Capodimonte, Via Moiariello 16, 80131 Napoli, Italy\label{aff6}
\and
Kapteyn Astronomical Institute, University of Groningen, PO Box 800, 9700 AV Groningen, The Netherlands\label{aff7}
\and
Leiden Observatory, Leiden University, Einsteinweg 55, 2333 CC Leiden, The Netherlands\label{aff8}
\and
Institute for Astronomy, University of Hawaii, 2680 Woodlawn Drive, Honolulu, HI 96822, USA\label{aff9}
\and
Institute for Astronomy, University of Edinburgh, Royal Observatory, Blackford Hill, Edinburgh EH9 3HJ, UK\label{aff10}
\and
Universit\'e de Strasbourg, CNRS, Observatoire astronomique de Strasbourg, UMR 7550, 67000 Strasbourg, France\label{aff11}
\and
Max-Planck-Institut f\"ur Astronomie, K\"onigstuhl 17, 69117 Heidelberg, Germany\label{aff12}
\and
Dunlap Institute for Astronomy \& Astrophysics, University of Toronto, 50 St. George Street, Toronto, ON M5S 3H4, Canada\label{aff13}
\and
Department of Physics and Astronomy, University of British Columbia, Vancouver, BC V6T 1Z1, Canada\label{aff14}
\and
School of Physics and Astronomy, University of Nottingham, University Park, Nottingham NG7 2RD, UK\label{aff15}
\and
INAF-Osservatorio di Astrofisica e Scienza dello Spazio di Bologna, Via Piero Gobetti 93/3, 40129 Bologna, Italy\label{aff16}
\and
ESAC/ESA, Camino Bajo del Castillo, s/n., Urb. Villafranca del Castillo, 28692 Villanueva de la Ca\~nada, Madrid, Spain\label{aff17}
\and
INAF-Osservatorio Astronomico di Brera, Via Brera 28, 20122 Milano, Italy\label{aff18}
\and
IFPU, Institute for Fundamental Physics of the Universe, via Beirut 2, 34151 Trieste, Italy\label{aff19}
\and
INAF-Osservatorio Astronomico di Trieste, Via G. B. Tiepolo 11, 34143 Trieste, Italy\label{aff20}
\and
INFN, Sezione di Trieste, Via Valerio 2, 34127 Trieste TS, Italy\label{aff21}
\and
SISSA, International School for Advanced Studies, Via Bonomea 265, 34136 Trieste TS, Italy\label{aff22}
\and
Dipartimento di Fisica e Astronomia, Universit\`a di Bologna, Via Gobetti 93/2, 40129 Bologna, Italy\label{aff23}
\and
INFN-Sezione di Bologna, Viale Berti Pichat 6/2, 40127 Bologna, Italy\label{aff24}
\and
Dipartimento di Fisica, Universit\`a di Genova, Via Dodecaneso 33, 16146, Genova, Italy\label{aff25}
\and
INFN-Sezione di Genova, Via Dodecaneso 33, 16146, Genova, Italy\label{aff26}
\and
Department of Physics "E. Pancini", University Federico II, Via Cinthia 6, 80126, Napoli, Italy\label{aff27}
\and
Dipartimento di Fisica, Universit\`a degli Studi di Torino, Via P. Giuria 1, 10125 Torino, Italy\label{aff28}
\and
INFN-Sezione di Torino, Via P. Giuria 1, 10125 Torino, Italy\label{aff29}
\and
INAF-Osservatorio Astrofisico di Torino, Via Osservatorio 20, 10025 Pino Torinese (TO), Italy\label{aff30}
\and
INAF-IASF Milano, Via Alfonso Corti 12, 20133 Milano, Italy\label{aff31}
\and
INAF-Osservatorio Astronomico di Roma, Via Frascati 33, 00078 Monteporzio Catone, Italy\label{aff32}
\and
INFN-Sezione di Roma, Piazzale Aldo Moro, 2 - c/o Dipartimento di Fisica, Edificio G. Marconi, 00185 Roma, Italy\label{aff33}
\and
Centro de Investigaciones Energ\'eticas, Medioambientales y Tecnol\'ogicas (CIEMAT), Avenida Complutense 40, 28040 Madrid, Spain\label{aff34}
\and
Port d'Informaci\'{o} Cient\'{i}fica, Campus UAB, C. Albareda s/n, 08193 Bellaterra (Barcelona), Spain\label{aff35}
\and
INFN section of Naples, Via Cinthia 6, 80126, Napoli, Italy\label{aff36}
\and
Dipartimento di Fisica e Astronomia "Augusto Righi" - Alma Mater Studiorum Universit\`a di Bologna, Viale Berti Pichat 6/2, 40127 Bologna, Italy\label{aff37}
\and
Jodrell Bank Centre for Astrophysics, Department of Physics and Astronomy, University of Manchester, Oxford Road, Manchester M13 9PL, UK\label{aff38}
\and
European Space Agency/ESRIN, Largo Galileo Galilei 1, 00044 Frascati, Roma, Italy\label{aff39}
\and
Universit\'e Claude Bernard Lyon 1, CNRS/IN2P3, IP2I Lyon, UMR 5822, Villeurbanne, F-69100, France\label{aff40}
\and
Institut de Ci\`{e}ncies del Cosmos (ICCUB), Universitat de Barcelona (IEEC-UB), Mart\'{i} i Franqu\`{e}s 1, 08028 Barcelona, Spain\label{aff41}
\and
Instituci\'o Catalana de Recerca i Estudis Avan\c{c}ats (ICREA), Passeig de Llu\'{\i}s Companys 23, 08010 Barcelona, Spain\label{aff42}
\and
Institut de Ciencies de l'Espai (IEEC-CSIC), Campus UAB, Carrer de Can Magrans, s/n Cerdanyola del Vall\'es, 08193 Barcelona, Spain\label{aff43}
\and
UCB Lyon 1, CNRS/IN2P3, IUF, IP2I Lyon, 4 rue Enrico Fermi, 69622 Villeurbanne, France\label{aff44}
\and
Mullard Space Science Laboratory, University College London, Holmbury St Mary, Dorking, Surrey RH5 6NT, UK\label{aff45}
\and
Department of Astronomy, University of Geneva, ch. d'Ecogia 16, 1290 Versoix, Switzerland\label{aff46}
\and
Universit\'e Paris-Saclay, CNRS, Institut d'astrophysique spatiale, 91405, Orsay, France\label{aff47}
\and
Max Planck Institute for Extraterrestrial Physics, Giessenbachstr. 1, 85748 Garching, Germany\label{aff48}
\and
Universit\"ats-Sternwarte M\"unchen, Fakult\"at f\"ur Physik, Ludwig-Maximilians-Universit\"at M\"unchen, Scheinerstr.~1, 81679 M\"unchen, Germany\label{aff49}
\and
INAF-Istituto di Astrofisica e Planetologia Spaziali, via del Fosso del Cavaliere, 100, 00100 Roma, Italy\label{aff50}
\and
INFN-Bologna, Via Irnerio 46, 40126 Bologna, Italy\label{aff51}
\and
School of Physics, HH Wills Physics Laboratory, University of Bristol, Tyndall Avenue, Bristol, BS8 1TL, UK\label{aff52}
\and
Aix-Marseille Universit\'e, CNRS/IN2P3, CPPM, Marseille, France\label{aff53}
\and
University Observatory, LMU Faculty of Physics, Scheinerstr.~1, 81679 Munich, Germany\label{aff54}
\and
INAF-Osservatorio Astronomico di Padova, Via dell'Osservatorio 5, 35122 Padova, Italy\label{aff55}
\and
Institute of Theoretical Astrophysics, University of Oslo, P.O. Box 1029 Blindern, 0315 Oslo, Norway\label{aff56}
\and
Jet Propulsion Laboratory, California Institute of Technology, 4800 Oak Grove Drive, Pasadena, CA, 91109, USA\label{aff57}
\and
Department of Physics, Lancaster University, Lancaster, LA1 4YB, UK\label{aff58}
\and
Felix Hormuth Engineering, Goethestr. 17, 69181 Leimen, Germany\label{aff59}
\and
Technical University of Denmark, Elektrovej 327, 2800 Kgs. Lyngby, Denmark\label{aff60}
\and
Cosmic Dawn Center (DAWN), Denmark\label{aff61}
\and
NASA Goddard Space Flight Center, Greenbelt, MD 20771, USA\label{aff62}
\and
Department of Physics and Astronomy, University College London, Gower Street, London WC1E 6BT, UK\label{aff63}
\and
Universit\'e de Gen\`eve, D\'epartement de Physique Th\'eorique and Centre for Astroparticle Physics, 24 quai Ernest-Ansermet, CH-1211 Gen\`eve 4, Switzerland\label{aff64}
\and
Department of Physics, P.O. Box 64, University of Helsinki, 00014 Helsinki, Finland\label{aff65}
\and
Helsinki Institute of Physics, Gustaf H{\"a}llstr{\"o}min katu 2, University of Helsinki, 00014 Helsinki, Finland\label{aff66}
\and
Laboratoire d'etude de l'Univers et des phenomenes eXtremes, Observatoire de Paris, Universit\'e PSL, Sorbonne Universit\'e, CNRS, 92190 Meudon, France\label{aff67}
\and
SKAO, Jodrell Bank, Lower Withington, Macclesfield SK11 9FT, UK\label{aff68}
\and
Centre de Calcul de l'IN2P3/CNRS, 21 avenue Pierre de Coubertin 69627 Villeurbanne Cedex, France\label{aff69}
\and
Universit\"at Bonn, Argelander-Institut f\"ur Astronomie, Auf dem H\"ugel 71, 53121 Bonn, Germany\label{aff70}
\and
Aix-Marseille Universit\'e, CNRS, CNES, LAM, Marseille, France\label{aff71}
\and
Dipartimento di Fisica e Astronomia "Augusto Righi" - Alma Mater Studiorum Universit\`a di Bologna, via Piero Gobetti 93/2, 40129 Bologna, Italy\label{aff72}
\and
Department of Physics, Institute for Computational Cosmology, Durham University, South Road, Durham, DH1 3LE, UK\label{aff73}
\and
Universit\'e Paris Cit\'e, CNRS, Astroparticule et Cosmologie, 75013 Paris, France\label{aff74}
\and
CNRS-UCB International Research Laboratory, Centre Pierre Bin\'etruy, IRL2007, CPB-IN2P3, Berkeley, USA\label{aff75}
\and
Institute of Physics, Laboratory of Astrophysics, Ecole Polytechnique F\'ed\'erale de Lausanne (EPFL), Observatoire de Sauverny, 1290 Versoix, Switzerland\label{aff76}
\and
Telespazio UK S.L. for European Space Agency (ESA), Camino bajo del Castillo, s/n, Urbanizacion Villafranca del Castillo, Villanueva de la Ca\~nada, 28692 Madrid, Spain\label{aff77}
\and
Institut de F\'{i}sica d'Altes Energies (IFAE), The Barcelona Institute of Science and Technology, Campus UAB, 08193 Bellaterra (Barcelona), Spain\label{aff78}
\and
European Space Agency/ESTEC, Keplerlaan 1, 2201 AZ Noordwijk, The Netherlands\label{aff79}
\and
Space Science Data Center, Italian Space Agency, via del Politecnico snc, 00133 Roma, Italy\label{aff80}
\and
Centre National d'Etudes Spatiales -- Centre spatial de Toulouse, 18 avenue Edouard Belin, 31401 Toulouse Cedex 9, France\label{aff81}
\and
Institute of Space Science, Str. Atomistilor, nr. 409 M\u{a}gurele, Ilfov, 077125, Romania\label{aff82}
\and
Dipartimento di Fisica e Astronomia "G. Galilei", Universit\`a di Padova, Via Marzolo 8, 35131 Padova, Italy\label{aff83}
\and
INFN-Padova, Via Marzolo 8, 35131 Padova, Italy\label{aff84}
\and
Caltech/IPAC, 1200 E. California Blvd., Pasadena, CA 91125, USA\label{aff85}
\and
Instituto de F\'isica Te\'orica UAM-CSIC, Campus de Cantoblanco, 28049 Madrid, Spain\label{aff86}
\and
Institut de Recherche en Astrophysique et Plan\'etologie (IRAP), Universit\'e de Toulouse, CNRS, UPS, CNES, 14 Av. Edouard Belin, 31400 Toulouse, France\label{aff87}
\and
Universit\'e St Joseph; Faculty of Sciences, Beirut, Lebanon\label{aff88}
\and
Departamento de F\'isica, FCFM, Universidad de Chile, Blanco Encalada 2008, Santiago, Chile\label{aff89}
\and
Universit\"at Innsbruck, Institut f\"ur Astro- und Teilchenphysik, Technikerstr. 25/8, 6020 Innsbruck, Austria\label{aff90}
\and
Department of Physics and Helsinki Institute of Physics, Gustaf H\"allstr\"omin katu 2, University of Helsinki, 00014 Helsinki, Finland\label{aff91}
\and
Infrared Processing and Analysis Center, California Institute of Technology, Pasadena, CA 91125, USA\label{aff92}
\and
Departamento de F\'isica, Faculdade de Ci\^encias, Universidade de Lisboa, Edif\'icio C8, Campo Grande, PT1749-016 Lisboa, Portugal\label{aff93}
\and
Instituto de Astrof\'isica e Ci\^encias do Espa\c{c}o, Faculdade de Ci\^encias, Universidade de Lisboa, Tapada da Ajuda, 1349-018 Lisboa, Portugal\label{aff94}
\and
Cosmic Dawn Center (DAWN)\label{aff95}
\and
Niels Bohr Institute, University of Copenhagen, Jagtvej 128, 2200 Copenhagen, Denmark\label{aff96}
\and
Universidad Polit\'ecnica de Cartagena, Departamento de Electr\'onica y Tecnolog\'ia de Computadoras,  Plaza del Hospital 1, 30202 Cartagena, Spain\label{aff97}
\and
European University of Technology EUt+, European Union\label{aff98}
\and
Institute of Space Sciences (ICE, CSIC), Campus UAB, Carrer de Can Magrans, s/n, 08193 Barcelona, Spain\label{aff99}
\and
Institut d'Estudis Espacials de Catalunya (IEEC),  Edifici RDIT, Campus UPC, 08860 Castelldefels, Barcelona, Spain\label{aff100}}

   \date{}

\abstract{An efficient and accurate separation of unresolved point sources from extended objects, usually referred to as star–galaxy separation, is essential for the scientific exploitation of deep wide-field imaging surveys, particularly at faint magnitudes where morphological differences become subtle.}{This paper presents a detailed assessment of two morphology-based classification methods applied to single-band, ground-based imaging from the Ultraviolet Near-Infrared Optical Northern Survey (UNIONS), using \textit{Euclid} observations as a high-fidelity reference to assess their performance, and to train one of the methods.}{We analysed $8$~deg$^2$ of overlapping UNIONS $r$-band and \textit{Euclid} \IE data, comprising more than $7\times10^{5}$ sources. The \textit{Euclid} data enable robust morphological labelling down to \IE$\simeq 26$, which we used to train and validate two classification approaches applied to UNIONS alone; i.e. a curve-of-growth method and a supervised neural-network classifier, both of which are based on aperture photometry and image-quality information.}{The curve-of-growth method yields an almost perfectly complete point-source sample ($>95\%$), but its purity decreases rapidly at magnitudes fainter than $r \simeq 22.5$. The neural-network approach significantly improves the purity at the faint end, maintaining $\geq 80\%$ purity down to $r \simeq 24$, while the completeness decreases to almost $50\%$ at similar depths. In both cases, the classification performance strongly depends on the image quality of the ground-based data.}{This work demonstrates how \textit{Euclid} data can be used to develop, calibrate, and validate star–galaxy separation methods for ground-based surveys, which are particularly useful in regions not covered by \textit{Euclid}. Both methods have already been applied to UNIONS data and are readily applicable to future surveys such as LSST, providing flexible solutions depending on whether completeness or purity is prioritised.}

\titlerunning{UNIONS–Euclid point source--galaxy separation}
\authorrunning{Thomas et al.}

   \maketitle
\nolinenumbers

\section{Introduction}

For many studies in modern astronomy, it is crucial to correctly identify and classify unresolved point-like sources, such as stars or quasi-stellar objects (QSOs), from extended objects, i.e., resolved galaxies. This process is commonly, though somewhat imprecisely, referred to as star–galaxy separation. This distinction underpins a wide range of research areas, from mapping the morphology of the Milky Way outskirts \citep[e.g.][]{fukushima_2018, fukushima_2019, thomas_2018, starkenburg_2019, amarante_2024,medina_2024} and detecting ultra-faint dwarf galaxies in the Local Group \citep[e.g.][]{martin_2013, drlica-wagner_2015, smith_2024} to probing the distribution of dark matter on cosmological scales through the weak-lensing distortion of distant galaxies \citep{ross_2011, soumagnac_2015, guerrini_2025, mpetha_2025}. Misclassifications between these two classes of objects can therefore introduce significant biases and lead to erroneous scientific conclusions.

At bright magnitudes, distinguishing stars from galaxies is relatively 
straightforward; the former outnumber the latter, which are, moreover, very 
extended sources on the sky. However, as imaging surveys reach fainter 
magnitudes, this separation becomes increasingly challenging 
\citep[e.g.][]{slater_2020, zhao_2026}. Indeed, at fainter magnitudes the 
fraction of stars decreases while the apparent angular size of resolved 
galaxies also decreases, making them harder to disentangle from point 
sources. For instance, using SDSS Stripe 82, \citet{slater_2020} showed that stars outnumber galaxies down to $r \simeq 20.8$, where the two populations are equal in number. Beyond this limit, the star-to-galaxy ratio drops 
exponentially, reaching approximately 1:15 at $r \sim 24$, highlighting 
the increasing difficulty of reliable star--galaxy separation at the depths 
reached by modern wide-field surveys. For this reason, the separation between point sources and extended objects is already difficult in the faintest regime of existing surveys such as the Sloan Digital Sky Survey \citep[SDSS;][]{york_2000,vasconcellos_2011}, Pan-STARRS \citep{chambers_2016}, or the Dark Energy Survey \citep[DES;][]{farrow_2014, tachibana_2018, sevilla-noarbe_2018}. This challenge will become even more critical for the new generation of wide and deep surveys, such as the Ultraviolet Near-Infrared Optical Northern Survey \citep[UNIONS;][]{gwyn_2025}, the Legacy Survey of Space and Time \citep[LSST;][]{ivezic_2019}, or the \textit{Euclid} mission \citep{Scaramella-EP1,EuclidSkyOverview}, which will reach magnitudes where galaxies largely outnumber stars \citep{lupton_2001, fadely_2012, sevilla-noarbe_2018, slater_2020}. 

Morphological approaches remain the cornerstone of point-source versus extended-object separation. Early studies based on photographic plates already exploited image moments and surface-brightness profiles to distinguish unresolved from extended sources \citep[e.g.][]{macgillivray_1976, kron_1980, heydon-dumbleton_1989, yee_1991}. With the advent of CCD detectors, automated pipelines such as \textsc{SExtractor} \citep{bertin_1996} enabled efficient star–galaxy separation using morphological parameters including the flux radius or the stellarity index. These classical approaches typically rely on comparing flux measurements obtained with different methods such as point-spread-function (PSF) fitting photometry versus aperture-based estimates including Kron or Petrosian magnitudes \citep{petrosian_1976, kron_1980}, or on specific metrics such as the difference between PSF and model fluxes, as implemented in the SDSS \citep{lupton_2001}, Pan-STARRS \citep{farrow_2014}, and Hyper Suprime-Cam Subaru Strategic Program \citep[HSC-SSP;][]{bosch_2018} large surveys.

In recent years, more sophisticated methods have been developed that exploit both morphological and colour information \citep[e.g.][]{hildebrandt_2012,tachibana_2018,sevilla-noarbe_2018} and inject artificial objects into real data (e.g. \citealt{urbano_2025}; Hui et al. in prep.). Combined with Bayesian inference techniques \citep[e.g.][]{fadely_2012} and supervised or unsupervised machine-learning algorithms such as deep neural networks \citep{breiman_2001, gatto_2026} and boosted decision trees, these methods have been successfully applied to improve classification accuracy at faint magnitudes \citep[e.g.][]{clarke_2020, cunha_2022, cook_2024}. For example, this is used in the \textit{Euclid} mission \citep{tucci_2025}, where a combination of supervised machine-learning classifiers and SED-fitting techniques is applied to multi-band photometry from both Euclid and ground-based surveys to separate stars, galaxies, and QSOs. While these approaches can outperform purely morphological classifiers \citep[see][]{robin_2007, gatto_2026}, their requirement of multi-band photometry limits their applicability in the faintest regimes, where some bands do not reach sufficient depth. In addition, methods based on multi-band photometry may struggle to correctly identify rare or variable sources such as RR Lyrae stars or transient objects.

In this paper, we present two new methods to separate point sources from extended objects using the $r$-band photometry from UNIONS. The approaches introduced here are directly applicable to upcoming large-scale surveys such as LSST, whose image quality and depth are comparable to those of the UNIONS $r$-band data. 

The UNIONS dataset used in this analysis is described in Sect.~\ref{sec:data}. Both methods leverage high-quality morphological labels derived from the \textit{Euclid} survey, as detailed in Sect.~\ref{sec:euclidLabel}. The methodologies are presented in Sects.~\ref{sec:method1} and~\ref{sec:method2}. Finally, we summarise our findings and conclusions in Sect.~\ref{sec:Conclusion}.

\section{Data} \label{sec:data}

\begin{figure}
\includegraphics[angle=0,clip,width=8.5cm]{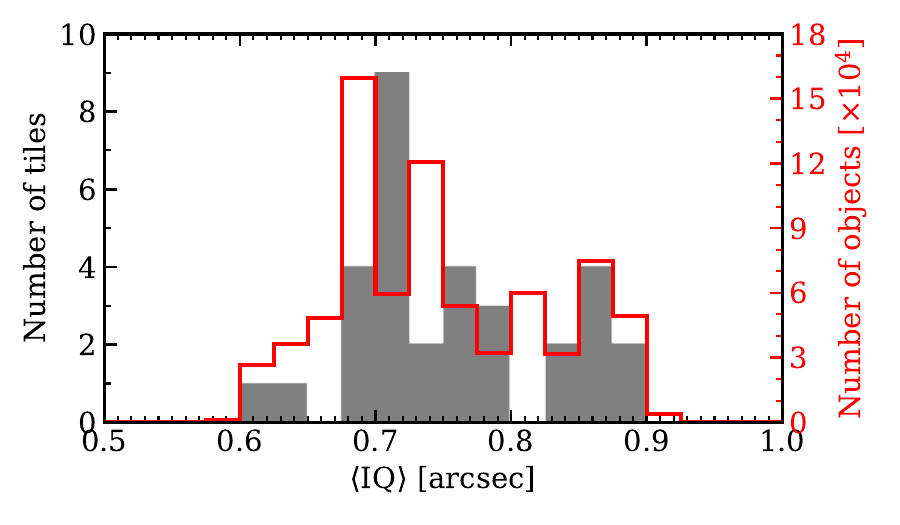}
   \caption{Distribution of the average image quality of the 32 tiles (plain grey histogram) and that of the different individual sources (red step histogram) used in this work. The image quality is measured as the FWHM of point sources in the UNIONS $r$ band.} 
\label{fig:IQ}
\end{figure}

The data used in this study come from an overlapping region observed both by the $r$ band of the Canada–France Imaging Survey (CFIS) carried out at the Canada-France-Hawai'i Telescope (CFHT), which is part of the UNIONS collaboration,\footnote{For simplicity, we refer to this dataset as the UNIONS $r$-band data in the rest of the paper.} and by the \textit{Euclid} space telescope. The analysed area consists of 32 tiles of $0.5 \times 0.5$~deg$^2$ each. Among these, 28 form a contiguous region covering approximately 7~deg$^2$ centred at (RA, Dec) = (17h\,16m, +59\degr), while the remaining four tiles are located in a separate region. The average image quality (IQ) of the UNIONS $r$-band photometry --- measured as the full width at half maximum (FWHM) of point sources --- of each of these 32 tiles is shown in Fig.~\ref{fig:IQ}. It ranges from 0.6 to 0.9~arcsec, with a median of 0.74 arcsec. The high image quality for a ground-based survey, the homogeneity of the wide coverage, and the fact that this is the band used to perform forced photometry in the UNIONS survey justified our choice to use the $r$ band in this work.

The UNIONS $r$ band was observed using MegaCam on CFHT. The individual images were calibrated using Gaia DR3 as an astrometric reference, and Pan-STARRS photometry was converted into the MegaCam filter system as a reference. The individual images were resampled and stacked onto a set of tiles covering the sky. The stacked images have a pixel scale of 0.1857~arcsec, which slightly oversamples the native MegaCam pixel scale. Further details are available in Section 3.1 of \citet{gwyn_2025}. 

For \textit{Euclid} data, the starting point is the stacked images produced by the pipeline used in the Euclid Wide Survey to process the visible instrument (VIS) data \citep{EuclidSkyVIS,Q1-TP002} in the \IE photometric band for Data Release~1 \citep{DR1cite}. Similar to the UNIONS data, the images were calibrated and stacked on a set of tiles, although the tiling scheme is different. SExtractor \citep{bertin_1996} is run on the \IE tiles to produce a catalogue containing the small aperture photometry required for the star--galaxy separation.

We only selected sources detected in both the UNIONS $r$ band and the \textit{Euclid} \IE band, with valid photometry (i.e. with generous selection of a measured magnitude\footnote{All magnitudes in this paper are in the AB system.} between 0 and 30 in both surveys) and valid aperture-magnitude measurements at all diameters used in this work (using a broad criterion on the photometric uncertainties: $\delta\mathrm{mag_{ap,i}} < 1.0$). The dataset was further cleaned to include only objects with measured isophotal semi-major and semi-minor axes in the UNIONS $r$ band. The final sample consists of $756\,802$ objects. We note that both the \textit{Euclid} and UNIONS catalogues are incomplete in regions of high surface brightness, as sources projected onto bright galaxy backgrounds are often not detected or fail the photometric quality cuts. As a result, the present analysis primarily focuses on relatively isolated sources, with blended or overlapping systems being largely excluded from the sample.

\begin{figure}
\includegraphics[angle=0,clip,width=8.5cm]{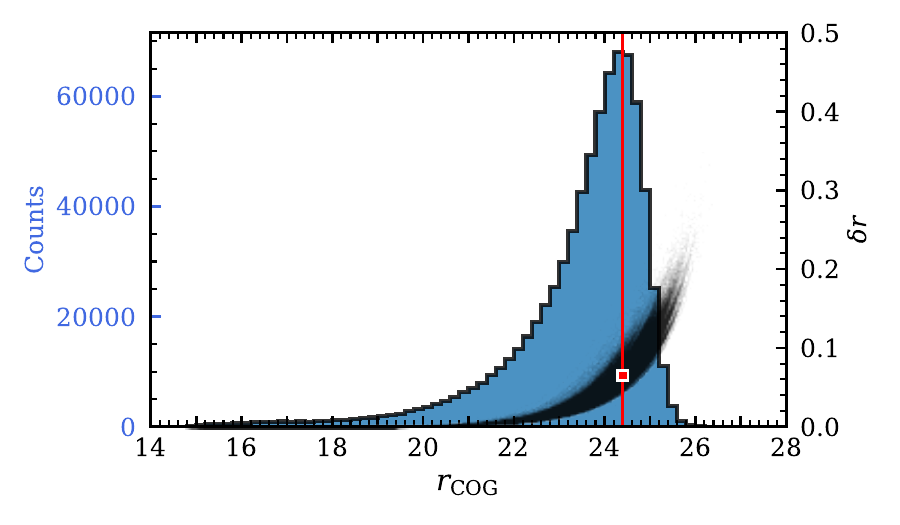}
   \caption{Number counts of the apparent magnitudes in terms of $r_\mathrm{COG}$ for all sources used in this analysis, shown as a blue histogram. Superimposed black markers show the magnitude uncertainties as a function of $r_\mathrm{COG}$. The vertical red line and square indicate the turnover in the luminosity function (i.e. the point of severe source loss). This occurs at $r_\mathrm{COG} \sim 24.4$, which has a corresponding median magnitude uncertainty of 0.065~mag or a detection of roughly 8.3$\sigma$.} 
\label{fig:LF}
\end{figure}

We show the distribution of this sample as a function of the UNIONS $r$-band curve-of-growth magnitudes ($r_\mathrm{COG}$; see Appendix~\ref{app:rcog}) and the magnitude uncertainty as a function of apparent magnitude in Fig.~\ref{fig:LF}. The luminosity function turnover indicates the apparent magnitude at which source loss becomes severe. This occurs at $r_\mathrm{COG} = 24.4$~, and there is a median error of 0.065~mag (corresponding to a signal-to-noise ratio of roughly 8.3), which is 0.6 magnitudes brighter than the 5$\sigma$ point-source depth in $r_\mathrm{COG} =25.0$).

\begin{figure}
\includegraphics[angle=0,clip,width=8.5cm]{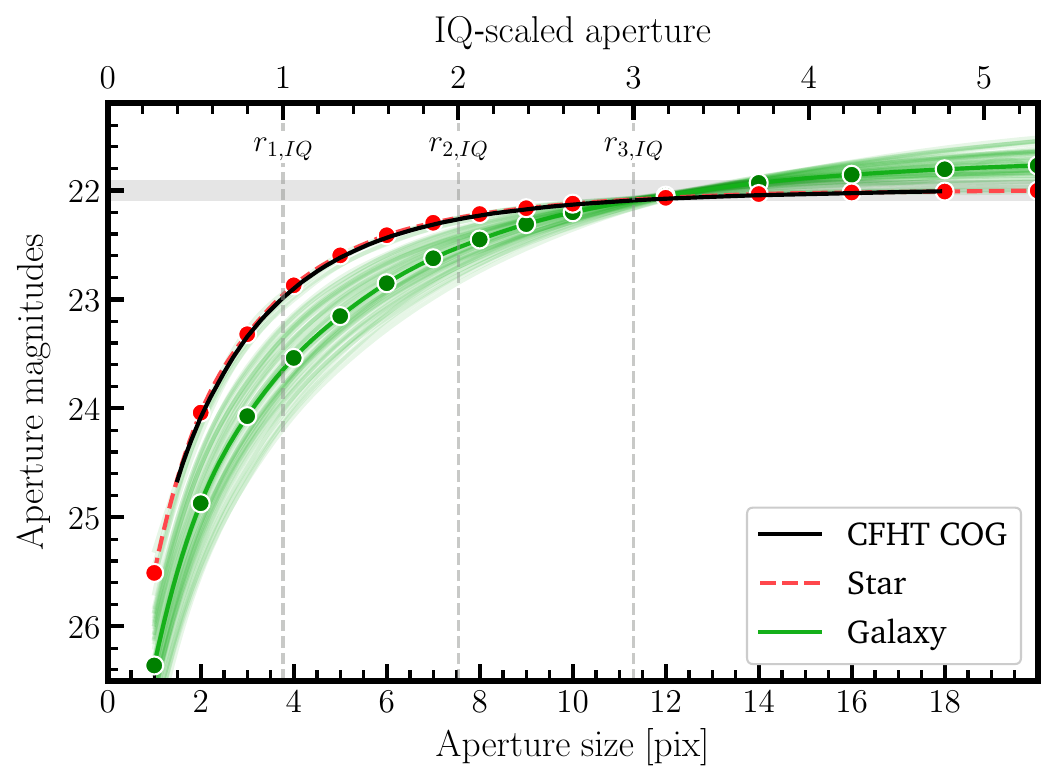}
\caption{Comparison between the curves of growth of a typical point source (dashed red line and circles) and those of 56 individual galaxies (dashed green lines and circles) with the same image quality ($0.7~\mathrm{arcsec}$) and similar apparent magnitude ($r_\mathrm{COG} \sim 22$, indicated by the grey shaded region). The solid green line shows the median curve of the galaxy population. The lower $x$-axis gives the aperture magnitude as a function of aperture size in pixels, while the upper $x$-axis expresses the aperture size in units of the image quality. The vertical dashed grey lines indicate the apertures at which the magnitude is interpolated for use in Method~1 (Sect.~\ref{sec:method1}). The solid black line shows the curve of growth of the empirically derived CFHT point spread function.}
\label{fig:COG}
\end{figure}

\begin{figure}
\includegraphics[angle=0,clip,width=9cm]{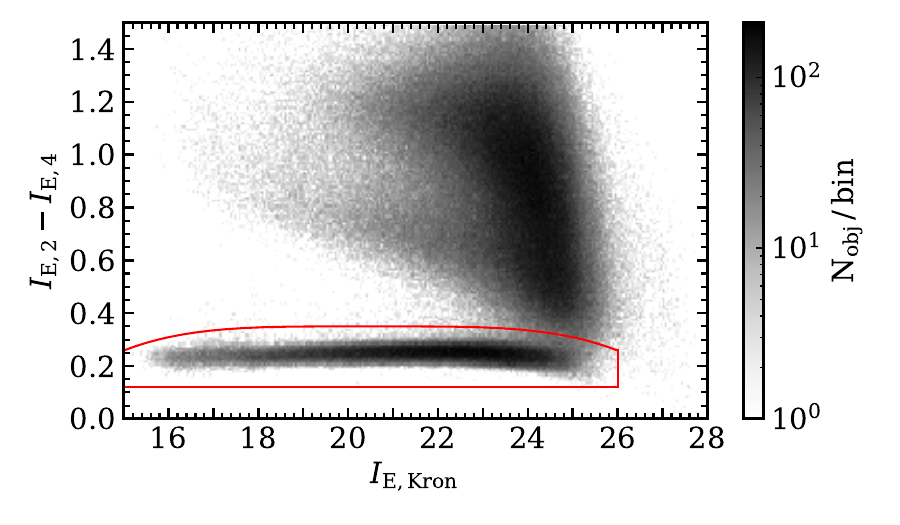}
   \caption{Difference between the \textit{Euclid} visible-band aperture magnitudes at 2 and 4 pixels as a function of the Kron-like elliptical aperture magnitude in the same band. The red contour shows the separation between point and extended sources.} 
\label{fig:sep_euclid}
\end{figure}

\section{Point versus extended source separation} \label{sec:method}

A traditional approach to distinguishing unresolved point-like sources, such as stars or QSOs, from extended objects, such as resolved galaxies, relies on their morphological properties. In particular, this involves comparing the magnitude measured within different aperture sizes centred on the source, for example by comparing the PSF magnitude with the Kron or Petrosian magnitude \citep{petrosian_1976, kron_1980}.

For point sources, the flux distribution closely follows the PSF, which, for ground-based observations such as UNIONS, can be approximated to the first order by a Moffat profile \citep{moffat_1969} with an IQ comparable to the PSF size, as illustrated in Fig.~\ref{fig:COG}. The empirical curve of growth for MegaCam was computed over $100\,000$ images drawn from the entire MegaCam archive. Flux is measured in a series of closely spaced apertures. Scaling the resulting curves of growth by the IQ of each image results in an extremely tight relation with a scatter of a few mmag. The relation is insensitive to the filter and does not show any dependency on the ellipticity of the PSF, except in cases of lost tracking. In effect, the one-dimensional radially averaged PSF can be described as a simple function of IQ.
For \textit{Euclid} VIS photometry, \citet{cuillandre_2025} showed that the PSF is nearly diffraction-limited, closely matching a pure diffraction halo; its wings follow the characteristic Bessel-function profile of an Airy disc, with an approximate $r^{-3}$ slope. Beyond a diameter of $3$~IQ, the measured magnitude remains nearly constant, since more than 99.9\% of the total flux of a point source is enclosed within this diameter. In contrast, extended sources such as galaxies exhibit a broader spatial flux distribution.

In the following sections, we present two different approaches for morphologically separating point sources from extended objects in the UNIONS surveys. The training and performance assessment of these methods were carried out using a high-fidelity reference sample derived from the \textit{Euclid} data, as described below.

\subsection{The \textit{Euclid} reference sample} \label{sec:euclidLabel}

\begin{figure}
\includegraphics[angle=0,clip,width=9cm]{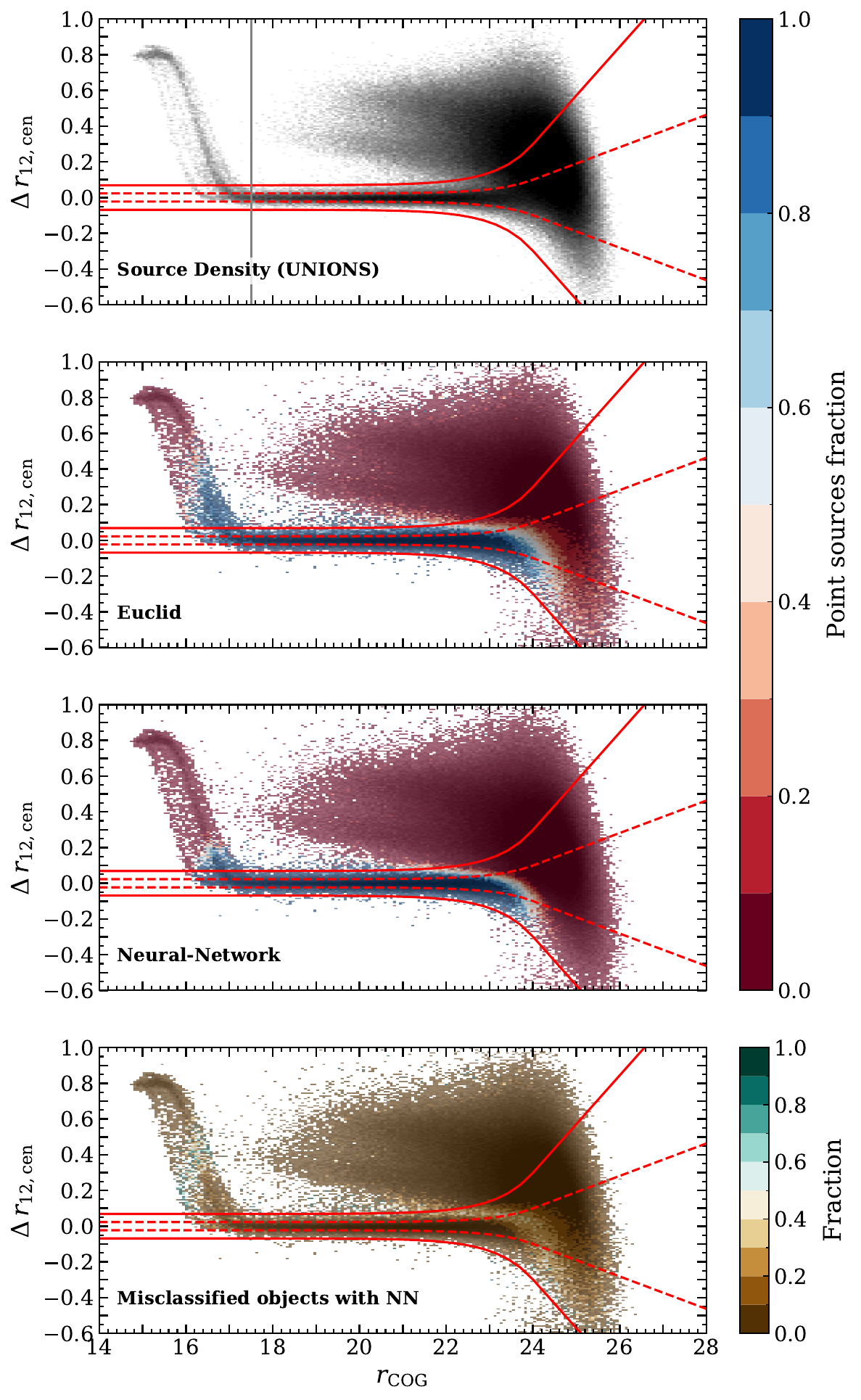}
   \caption{Re-centred UNIONS $r$-band aperture magnitude difference measured at 1\,$\times$ and 2\,$\times$ IQ ($\Delta r_{12,\,\mathrm{cen}}$), as a function of $r_\mathrm{COG}$. The {\it upper panel} shows the source density, with the point-source locus centred at $\Delta r_{12,\,\mathrm{cen}} = 0$ by construction. Severe saturation is observed at magnitudes brighter than $r_\mathrm{COG} = 17.5$ (vertical grey line). The {\it middle panels} show the same quantity, colour-coded by the average fraction of point sources per bin, according to the \textit{Euclid}-based reference classification (Sect.~\ref{sec:euclidLabel}) and the neural-network classification (Sect.~\ref{sec:method2}), respectively. The {\it lower panels} show the fraction of misclassified objects of the neural-network compared to the \textit{Euclid}-based reference. In each panel, the dashed red lines show how the $1\sigma$ width of the point-source locus increases as a function of magnitude down to $r_\mathrm{COG} = 24$, after which it is extended linearly. The solid red line shows the $3\sigma$ boundary, which defines the nominal selection criterion for point sources. }

\label{fig:sep_cfis}
\end{figure}

Thanks to the exquisite image quality of the \textit{Euclid} data \citep{cuillandre_2025,EuclidSkyOverview}, nearly constant throughout its entire field of view, it is possible to reliably distinguish point sources from extended objects morphologically, even at magnitudes as faint as $I_{\mathrm{E},\,\text{Kron}} = 25.5$, as shown in Fig.~\ref{fig:sep_euclid}. In this figure, point sources occupy a well-defined region described by
\begin{equation}
\label{eq:mag}
\begin{cases}
0.12 \leq I_\mathrm{E,2} - I_\mathrm{E,4} \leq -0.0001 \  (I_{\mathrm{E},\,\text{Kron}} - 20.5)^4 + 0.35, \\
I_{\mathrm{E},\,\text{Kron}} \leq 26.0,
\end{cases}
\end{equation}
where $I_{\mathrm{E},j}$ denotes the aperture magnitude measured within a diameter of $j$ pixels of 0.1~arcsec, and $I_{\mathrm{E},\,\text{Kron}}$ is the Kron elliptical aperture magnitude in the \IE band.  The cut-off value of $I_{\mathrm{E},\,\text{Kron}} \leq 26.0$ was selected to maximise the completeness of the point-source sample, at the cost of higher contamination. Changing this value to $I_{\mathrm{E},\,\text{Kron}} \leq 25.5$ does not change the conclusions drawn in this paper. With this selection in place, out of the $756\,802$ objects in the initial dataset $127\,612$ are classified as point sources, while the other $629\,190$ are identified as resolved extended objects. Cross-matching this sample with the data from the third data release of \textit{Gaia} \citep[DR~3][]{gaia_DR3}, $3.4\%$ of the $37\,897$ point sources in common are classified as QSOs according to the \textit{Gaia} classification probability \textsc{classprob\_dsc\_combmod\_quasar}$>0.5$.

In the remainder of this paper, we adopt this clear morphological separation as the ground-truth classification (labels). This \textit{Euclid}-based reference sample is used to evaluate the performance of the different point-source versus extended-object separation methods applied to the UNIONS photometry. Note that despite the exquisite \textit{Euclid} photometry, it is possible that contamination from compact galaxies or non-fully resolved star clusters is present in this selection, in particular from those further than $\sim 20$ Mpc \citep{EROFornaxGCs,EP-Voggel}.  However, at this magnitude depth, such contamination is marginal. Indeed, \citet{EP-Voggel} estimated that $\sim 350\,000$ globular clusters will be detectable across the $14\,000$~deg$^2$ of the Euclid Wide Survey, of which $\sim 115\,000$ will have high-quality photometry ($\IE \leq 24.5$). This translates to an upper limit of $\simeq 16$ globular clusters per deg$^2$ that could be identified as point sources, corresponding to at most $\sim 160$ objects in the area considered here -- i.e. far fewer than the number of objects we considered as point sources, even at the faintest magnitudes.

\begin{figure*}
\includegraphics[angle=0,clip,width=18cm]{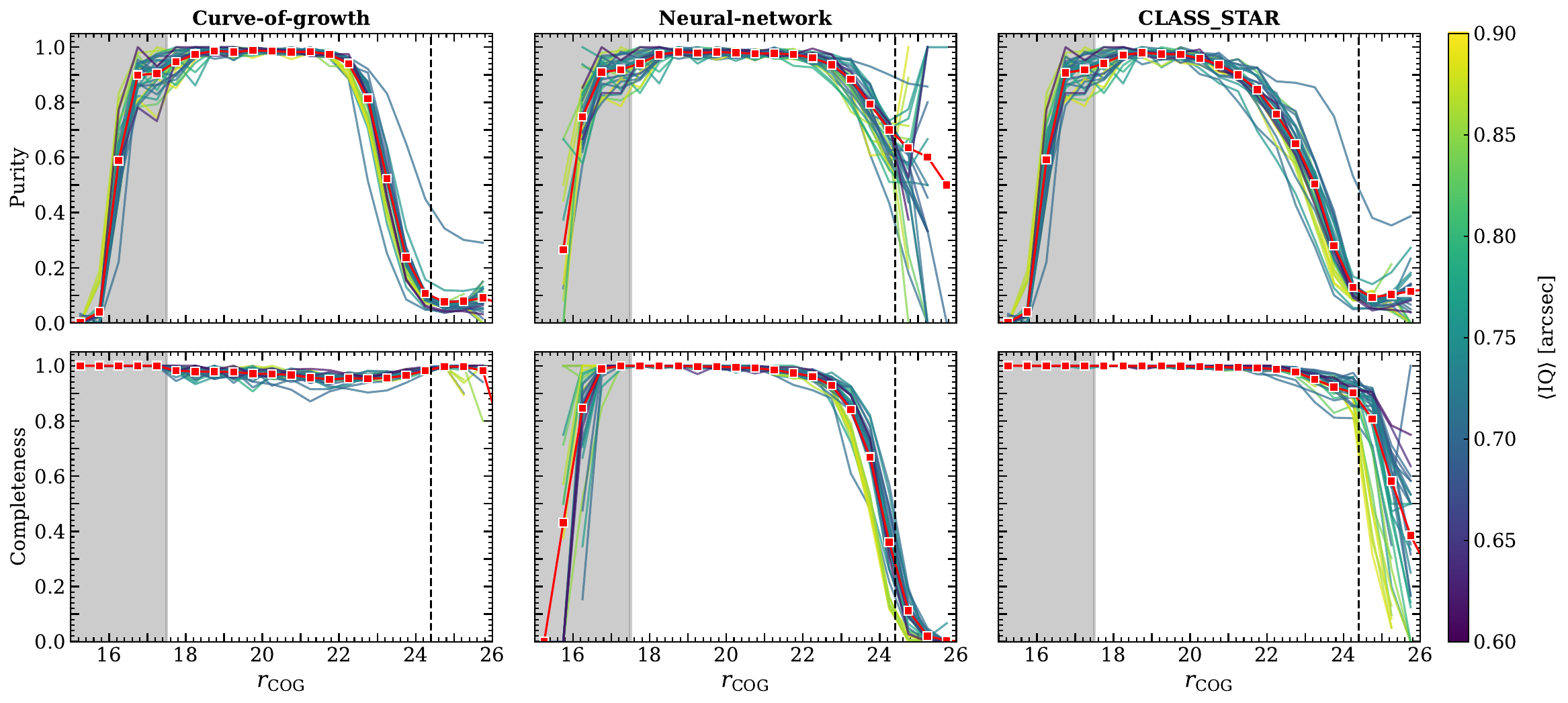}
   \caption{Purity and completeness of the point-source classification as a function of magnitude, in bins of 0.5~mag, for the curve-of-growth method ({\it left panels}), the neural-network method ({\it middle panels}), and the \textsc{SExtractor} stellarity index ($\textsc{class\_star} > 0.5$, {\it right panels}). Each line shows the evolution of these quantities for individual tiles, colour-coded by their average IQ. The red lines show the global evolution of purity and completeness for the entire dataset. The grey region indicates the range where the UNIONS $r$-band data become saturated, and the vertical dashed lines indicate the location of the turnover in the luminosity function ($r_\mathrm{COG} = 24.4$).} 
\label{fig:pur_comp_method}
\end{figure*}

\begin{figure}
\includegraphics[angle=0,clip,width=8.5cm]{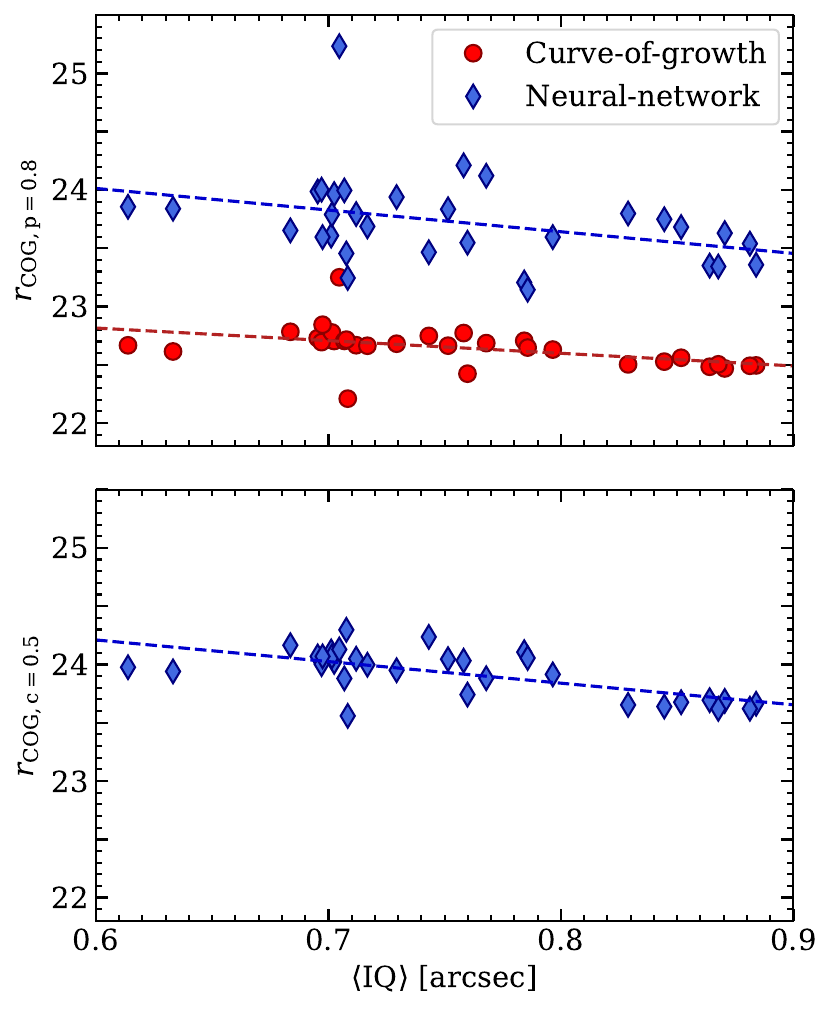}
\caption{{\it Top panel} shows the variation of the magnitude at which the purity reaches 0.8 as a function of the average IQ of the UNIONS $r$-band tiles for both methods. The {\it bottom panel} shows the variation of the magnitude at which the completeness reaches 0.5 as a function of the IQ. The first method is not shown in the bottom panel, since its completeness remains approximately constant down to $r_\mathrm{COG} = 26.0$.}

\label{fig:lim_put_comp}
\end{figure}

\subsection{Method 1: Morphological curve-of-growth method} \label{sec:method1}

The first method (referred to as the curve-of-growth method) for point-source identification used for the ground-based UNIONS survey is performed morphologically in the co-added $r$-band images, which were observed with the MegaCam imager at CFHT, and was initially developed as part of a study to detect Local Group ultra-faint dwarf galaxies \citep{smith_2024}. Aperture photometry is measured in 20 concentric apertures with diameters of 1, 2, 3, 4, 5, 6, 7, 8, 9, 10, 12, 14, 16, 18, 20, 22, 25, 30, 35, and 40 pixels on the co-adds, and the IQ is recorded for each source. We interpolated between all aperture magnitudes using a cubic spline and recorded the magnitude at precisely 1~IQ ($r_{1, \mathrm{IQ}}$) and 2~IQ ($r_{2, \mathrm{IQ}}$). In Fig.~\ref{fig:COG}, we demonstrate this process by showing the concentric aperture magnitudes and the interpolated curve-of-growth for both a star and a galaxy (as identified in the {\it Euclid} reference sample) with the same $r_\mathrm{COG} = 22$ and IQ (0.7~arcsec).

By taking $r_{1,\,\mathrm{IQ}} - r_{2,\,\mathrm{IQ}}$, which corresponds to a flux ratio between the inner and outer apertures, a point source locus is visible, with some scatter about it. The median value of this locus is $r_{1,\,\mathrm{IQ}} - r_{2,\,\mathrm{IQ}} = 0.7068$, although we find that when the set of images contributing to a co-add has widely varying IQs, the locus position can shift by up to 0.06~mag from this fiducial value. We therefore computed the locus position per tile and defined the re-centred aperture magnitude difference as follows:
\begin{equation}
    \Delta r_{12,\,\mathrm{cen}} \equiv r_{1,\,\mathrm{IQ}} - r_{2,\,\mathrm{IQ}} - 
    \widetilde{r}_{12},
\end{equation}
\noindent where $\widetilde{r}_{12}$ is the per-tile median of the point source locus, such that the locus is placed exactly at zero for every tile, ensuring that no region of the sky experiences catastrophic source loss should the locus shift towards the edge of the fiducial selection boundary. This is shown in the upper panel of Fig.~\ref{fig:sep_cfis}, where three distinct populations can be identified. The point-source locus is located around $\Delta r_{12,\,\mathrm{cen}} = 0$. Extended sources form a poorly defined cloud at larger values of $\Delta r_{12,\,\mathrm{cen}}$, which merges with the point source locus at fainter magnitudes ($r_\mathrm{COG} \simeq 24.4$) as photometric uncertainties increase. Saturated point sources form a vertical feature at magnitudes brighter than $r_\mathrm{COG} = 17.5$~mag and are therefore excluded from the remainder of the analysis. They account for $\sim 1\%$ of the initial dataset. For comparison, the middle panel shows the same quantities colour-coded by the fraction of point sources identified in the \textit{Euclid}-based reference sample (see Sect.~\ref{sec:euclidLabel}), whilst the lower panel shows the same quantity for point sources identified with the neural network method (see Sect.~\ref{sec:method2}).

To develop robust criteria for selecting point sources, we characterised the point-source locus and quantified the growth in its width as a function of magnitude, mostly driven by the increase of the photometric uncertainties. To avoid contamination from the cloud of extended sources, we took all sources with 
$\Delta\,r_{12,\mathrm{cen}} < 0$ and measured the 68th percentile 
(i.e.\ the $1\sigma$ half width of the locus) for $17.5 \leq r_\mathrm{COG} \leq 23$. We fitted an exponential curve to these measurements, which was used to define the growth in the locus width down to $r_\mathrm{COG} = 24$, after which the curve was extended linearly as a tangent to the exponential curve to moderate its growth. The piece-wise exponential and linear curve along the bottom half of the point source locus is defined by
\begin{equation}
\label{eq:2}
\begin{aligned}
\begin{cases}
-2.818\times 10^{-14}\, \exp(1.193~r_\mathrm{COG}) - 0.0229, & r_\mathrm{COG} \leq 24,\\
-9.093\times 10^{-2}~r_\mathrm{COG} +2.0828, & r_\mathrm{COG} > 24,
\end{cases}
\end{aligned}
\end{equation}
\noindent which was then symmetrically reflected across $\Delta\,r_{12,\mathrm{cen}} = 0$ to define the 1$\sigma$ width of the point-source locus. Every source was then assigned a distance from the median locus, scaled to the width of the bounding curves at its magnitude, which we refer to as $s_{21}$.

We define the selection criterion for point sources as $|s_{21}| \le 3$, which is shown as solid red lines in the upper panel of Fig.~\ref{fig:sep_cfis}; there, we show a point-source catalogue 
with excellent completeness but sacrificing purity. The results of morphological classification from the curve-of-growth method are shown for the entire dataset, as well as for each tile individually, in the left panels of Fig.~\ref{fig:pur_comp_method}. The purity of the sample exceeds 90\% for $r_\mathrm{COG} \le 22.4$, reaches 50\% at $r_\mathrm{COG} \simeq 23.3$, and falls below 10\% for $r_\mathrm{COG} > 24.3$. The completeness remains above 95\% across the entire range of non-saturated magnitudes 
($r_\mathrm{COG} > 17.5$).

For comparison, the right panels of Fig.~\ref{fig:pur_comp_method} show point-source selection based on the \textsc{SExtractor} stellarity index ($\textsc{class\_star} > 0.5$). In that case, the purity falls below 80\% at $r_\mathrm{COG} \simeq 22.0$ and below 50\% at $r_\mathrm{COG} \simeq 23.3$, whilst the completeness only remains above 95\% for $r_\mathrm{COG} \le 23.3$. The curve-of-growth method represents a clear improvement compared to this approach, particularly in terms of purity at intermediate magnitudes and completeness at the faint end.

We also tested more restrictive selection regions at the curves defined by $|s_{21}| \le 2$ and $|s_{21}| \le 1$, which leads to improvements in the 50\% purity level (occurring at 23.5 and 23.8, respectively) at the cost of completeness, which drops to 90\% and 69\%, respectively. This figure also shows that the purity at a given magnitude depends on the IQ of the tile. The purity decreases towards fainter magnitudes in a similar manner, following a sigmoidal function, for all tiles, but the decline begins at brighter magnitudes for tiles with poorer IQs. This trend is clearly visible in the upper panel of Fig.~\ref{fig:lim_put_comp}, which shows how the magnitude at which the purity reaches 0.8 ($r_\mathrm{COG,p=0.8}$) varies as a function of the IQ of the tile. For this method, we find that $r_\mathrm{COG,p=0.8}$ depends linearly on the IQ of the tile, with a slope of $-1.08$.  In contrast, as shown in Fig.~\ref{fig:pur_comp_method}, the completeness remains largely constant with respect to IQ. 

We also tested the separation capabilities when comparing the aperture magnitude at 1~IQ to an aperture magnitude at 3~IQ. This followed the same procedure as before, with the analogous $r_{13,\mathrm{cen}}$-shifted magnitude difference and $s_{31}$ parameter being defined. Adopting a selection criterion of $|s_{31}| \leq 3$, we find a similar completeness (within 1\% variation) across the range of non-saturated magnitudes at the percentage level, but the 90\% and 50\% purity levels occur 0.7 magnitudes shallower. Selecting in both $s_{21}$ and $s_{31}$ simultaneously gives identical results to the $s_{21}$ case alone. We conclude that a selection criterion of $|s_{21}| \leq 3$, derived solely from ground-based imaging, is sufficient to efficiently separate point sources from extended objects down to $r_\mathrm{COG} = 23.2$, which is 1.2~mag brighter than the turnover magnitude. At fainter magnitudes, this method yields a highly complete sample of point sources, albeit with an increasing level of contamination from extended objects.

\subsection{Method 2: Neural-Network morphological classification} \label{sec:method2}

The second method used to disentangle point-like sources from extended objects in the UNIONS survey relies on an artificial neural-network (ANN) classifier. The input features include the aperture photometry measured in 20 concentric apertures on the co-added $r$-band images described in Sect.~\ref{sec:method1}, together with the IQ measured at the source position and the isophotal semi-major and semi-minor axes measured with \textsc{SExtractor} on the co-added $r$-band images ($A_\mathrm{world}$ and $B_\mathrm{world}$). For each source, the ANN outputs the probability of being either a point source or an extended object.

The ANN adopts a simple architecture composed of four fully connected hidden layers of 64 neurons each. A rectified-linear-unit (ReLU) activation function is applied to each hidden layer to introduce non-linearity. The output layer is composed of two neurons, corresponding to the two classes (point-like and extended), followed by a soft-max activation function to produce probabilistic class assignments. The network is implemented using the \textsc{Keras} package \citep{chollet_2015,chollet_2018}.

The training was performed on the labelled data obtained from the \textit{Euclid} reference sample, with 80\% of the initial dataset used to train the network and the remaining 20\% used as a validation sample. The optimisation was performed with the Adam algorithm \citep{kingma_2014}, with successive learning rates of $10^{-3}$, $10^{-4}$, and $10^{-5}$, in order to refine convergence towards the minimum, using mini-batches of 2000 objects. The network was trained for a maximum of 500 epochs, with an early stopping criterion that halts training if the validation loss does not improve for 20 consecutive epochs. A dropout rate of 0.1 was applied between the last hidden layer and the output layer to reduce overfitting. The overall architecture and hyperparameters were determined by testing different configurations and selecting those yielding the best results on a randomly selected 10\% subset of the initial sample; these were removed from both the training and validation sets.

The model was trained to minimise the binary cross-entropy loss function, defined as
\begin{equation}
\mathcal{L} = - \left[ y_\mathrm{ext} \ln(1 - P_\mathrm{P.S.}) + y_\mathrm{P.S.} \ln(P_\mathrm{P.S.}) \right],
\label{eq:4}
\end{equation}
where $P_\mathrm{P.S.}$ is the probability assigned by the network for an object to be a point source and $y_\mathrm{P.S.}$ ($y_\mathrm{ext}$) is a binary indicator equal to one if the object is a point source (extended object) and zero otherwise.

Note that no weighting was applied to correct for the class imbalance of the sample between point sources ($\sim 17\%$) and extended objects ($\sim83\%$). Several weighting schemes were tested during training to try to correct for the imbalanced dataset at different magnitudes. However, the performance was lower than without weighting the samples, yielding a much lower purity at the expense of higher completeness, without surpassing the efficiency of the first method. We therefore opted to train the model without any sample weighting.

The performance of the ANN classifier is presented in the middle panels of Fig.~\ref{fig:pur_comp_method}, which show the purity and completeness of the point-source candidates, selected such that $P_\mathrm{P.S.}>0.5$, as a function of the magnitude, measured in bins of 0.5 mag. The purity is constant at 0.97 up to 22.4, and it then decreases slowly down to 0.74 at $r_\mathrm{COG} = 24.0$. The completeness remains close to one for sources brighter than $r_\mathrm{COG} \simeq 22.5$ and declines to 0.5 at $r_\mathrm{COG} = 24.1$. The relative importance of each input is further analysed in Appendix~\ref{app:feat_imp}. \footnote{A public version of the training algorithm is available at \url{https://github.com/GFThomas/StarGal_ground_Euclid}.}

Compared to the curve-of-growth method and the \textsc{SExtractor} stellarity index, the purity decreases more slowly with magnitude. This is clearly visible in the upper panel of Fig.~\ref{fig:lim_put_comp}, where a purity of 80\% is reached approximately one magnitude deeper than with the curve-of-growth method. However, this approach is more sensitive to the IQ of each tile; we find that $r_\mathrm{COG,p=0.8}$ depends linearly on IQ with a slope of $-1.86$, compared to $-1.08$ for the curve-of-growth method.

In contrast, the completeness is not constant but decreases following a sigmoidal profile as a function of magnitude, and it is strongly dependent on the IQ of the tile; this is illustrated in the lower panel of Fig.~\ref{fig:lim_put_comp}, where the magnitude at which a completeness of 50\% is reached ($r_\mathrm{COG,c=0.5}$) depends linearly on IQ with a slope of $-1.85$. Compared to the \textsc{SExtractor} stellarity index, this method is also less complete; completeness falls below 90\%, and the result is around 1.3~mag brighter ($r_\mathrm{COG} \simeq 22.9$ compared to $r_\mathrm{COG} \simeq 24.2$).

By adopting a more stringent \textsc{SExtractor} stellarity threshold of $\textsc{class\_star} > 0.85$, the completeness becomes comparable to that of the neural network method. However, the purity remains lower, with $r_\mathrm{COG,p=0.8} = 22.6$, 
 $1.1$~mag brighter than that achieved by the neural-network method. In conclusion, this method achieves a higher level of purity with regard to separating point sources from extended objects at faint magnitudes compared to the first method and to the \textsc{SExtractor} stellarity index, albeit at the cost of a less complete point-source sample.

\section{Conclusions} \label{sec:Conclusion}

This work presents and compares two independent, morphology-based methods for separating point sources from extended objects in the UNIONS survey, using the co-added $r$-band photometry. To this end, we used an $8\deg^2$ region overlapping with \textit{Euclid}, exploiting the exquisite quality of its PSF to label each object as point source or as an extended object. This classification was then used as the reference to analyse the performance of the two methods applied to UNIONS.

The first method, based on the curve-of-growth magnitude, provides a simple and robust morphological criterion that yields an almost perfectly complete sample of point sources across the entire non-saturated magnitude range. This method maintains high purity (above 90\%) for $r_\mathrm{COG} \le 22.5$, but drops rapidly to below 10\% at $r_\mathrm{COG} = 24.0$. However, its completeness remains above 95\% across the explored magnitude range. This approach is therefore particularly well suited for science cases that prioritise maximal completeness, even at the expense of increased contamination from extended objects at fainter magnitudes.

The second method employs a neural-network classifier trained on multiple morphological indicators, including the aperture flux profile, isophotal axes, and image quality. This machine-learning approach significantly improves the purity of the point source sample, staying above $\gtrsim 90\%$, with completeness above 90\% for $r_\mathrm{COG} \le 23.5$. Beyond this magnitude, it maintains 80\% purity for $r_\mathrm{COG} \lesssim 24$, while completeness declines rapidly, reaching 50\% at $r_\mathrm{COG} \simeq 24$.

For both methods, we find that their performance depends on the image quality of the co-added $r$-band tile. Based on our results, we conclude that the choice between the two methods should be guided by the scientific application. For instance, studies aimed at identifying ultra-faint dwarf (UFD) galaxies or stellar stream candidates, where maximising completeness is critical, may benefit from the curve-of-growth method. However, we caution that compact background galaxies may be misclassified as stars, and identified UFD candidates should subsequently be vetted against known galaxy-cluster catalogues or via multi-band photometry to confirm or reject individual detections -- a step that becomes particularly important at the depths of UNIONS and future LSST data, where exhaustive spectroscopic follow-up may not be feasible. In contrast, studies that require a very pure point-source sample, such as analyses of the morphology of the outskirts of the Milky Way, are better served by the neural-network approach.

Finally, both methods can be directly applied to future deep ground-based surveys such as LSST, which is planned to reach depths comparable to UNIONS in the first year, but with a median image quality of about $0.8$~arcsec. Under these conditions, the curve-of-growth method is expected to achieve a purity of around $80\%$ and a completeness above 95\% for $r_\mathrm{COG} \le 22.8$, while the neural-network method should maintain 80\% purity for $r_\mathrm{COG} \le 23.8$ and a completeness of around 70\%. However, observations obtained with the LSST ComCam show that the median image quality is worse, at $1.14$~arcsec across all bands \citep{LSST_2025}. In this case, a purity of $\sim 80\%$ is expected at $r_\mathrm{COG} \simeq 22.4$ for the curve-of-growth method and at $r_\mathrm{COG} \simeq 23.1$ for the neural-network method. These results demonstrate that both techniques provide reliable and complementary tools for future large-scale optical surveys, enabling the accurate separation of point sources and extended objects from single-band ground-based imaging.

 \begin{acknowledgements}
The authors thank Annie Robin for helpful discussions. GFT acknowledges support from the Agencia Estatal de Investigaci\'on del Ministerio de Ciencia e Innovaci\'on (AEI-MCIN) under grant number PID2023-150319NB-C21 and the grant RYC2024-051016-I funded by MCIN/AEI/10.13039/501100011033 and by the European Social Fund Plus. Co-funded by the European Union (Widening Participation, ExGal-Twin, GA 101158446). GB acknowledges support from the Agencia Estatal de Investigaci\'on del Ministerio de Ciencia e Innovaci\'on (AEI-MCIN) under grant number PID2023-150319NB-C21.
We are honored and grateful for the opportunity of observing the Universe from Maunakea and Haleakala, which both have cultural, historical and natural significance in Hawaii. This work is based on data obtained as part of the Canada-France Imaging Survey, a CFHT large program of the National Research Council of Canada and the French Centre National de la Recherche Scientifique. Based on observations obtained with MegaPrime/MegaCam, a joint project of CFHT and CEA Saclay, at the Canada-France-Hawaii Telescope (CFHT) which is operated by the National Research Council (NRC) of Canada, the Institut National des Science de l’Univers (INSU) of the Centre National de la Recherche Scientifique (CNRS) of France, and the University of Hawaii. This research used the facilities of the Canadian Astronomy Data Centre operated by the National Research Council of Canada with the support of the Canadian Space Agency. This research is based in part on data collected at Subaru Telescope, which is operated by the National Astronomical Observatory of Japan.
Pan-STARRS is a project of the Institute for Astronomy of the University of Hawaii, and is supported by the NASA SSO Near Earth Observation Program under grants 80NSSC18K0971, NNX14AM74G, NNX12AR65G, NNX13AQ47G, NNX08AR22G, 80NSSC21K1572 and by the State of Hawaii.
The authors acknowledge the use of the Canadian Advanced Network for Astronomy Research (CANFAR) Science Platform operated by the Canadian Astronomy Data Center (CADC) and the Digital Research Alliance of Canada (DRAC), with support from the National Research Council of Canada (NRC), the Canadian Space Agency (CSA), CANARIE, and the Canadian Foundation for Innovation (CFI). 
\AckDRone 
\AckEC
\end{acknowledgements}

\bibliographystyle{aa}
\bibliography{biblio}

\begin{appendix}
\section{UNIONS $r$-band curve-of-growth magnitude ($r_{\rm COG}$)}
\label{app:rcog}

The UNIONS $r$-band magnitudes used in this work are computed using a
curve-of-growth technique implemented in the \textsc{PTOUSE} (Parallel
Tessellated Optimal Uniform Source Extraction) pipeline. For each tile,
\textsc{SExtractor} is run with 20 closely spaced fixed apertures, and each
detected source is traced back to the specific set of input exposures that
contributed to it -- a region referred to as a \emph{patch}, an irregularly
shaped area of sky sharing the same list of input images, over which the
point-spread function (PSF) is homogeneous even though it can vary strongly
across a full tile.

Within each patch, the closely spaced apertures are used to map the curve of
growth, and the flux measured in the inner apertures is extrapolated to a
large reference aperture (equivalent to \textsc{SExtractor}'s
\texttt{MAG\_AUTO} for stellar sources, corresponding to roughly $5.15$
times the image quality/seeing). This extrapolation is performed at the
patch level because it is sensitive to the local IQ, which
is homogeneous within a patch. Because the CFIS/MegaCam radial PSF can be
described to good accuracy by a single curve scaled to the seeing, this
curve-of-growth extrapolation remains reliable even for patches with too
few stars to measure it directly.

The resulting curve-of-growth magnitude (denoted \texttt{MAG\_COG} in the
catalogs, and referred to here as $r_{\rm COG}$) provides a robust and
low-noise flux measurement for point sources, since it is derived from a
set of relatively small apertures and benefits from some resilience against
residual cosmic-ray contamination because multiple apertures are combined.
The corresponding $5\sigma$ point-source limiting magnitude
(\texttt{MAGLIM\_COG}) is likewise computed at the patch level, and can be
predicted from the depths of the input exposures when a given patch lacks
enough stars for a direct measurement. Further details on the curve-of-growth magnitudes are available online \footnote{\url{https://www.cadc-ccda.hia-iha.nrc-cnrc.gc.ca/en/community/unions/MegaPipe_CFIS_DR3.html}.}.

\begin{figure}
\includegraphics[angle=0,clip,width=8.0cm]{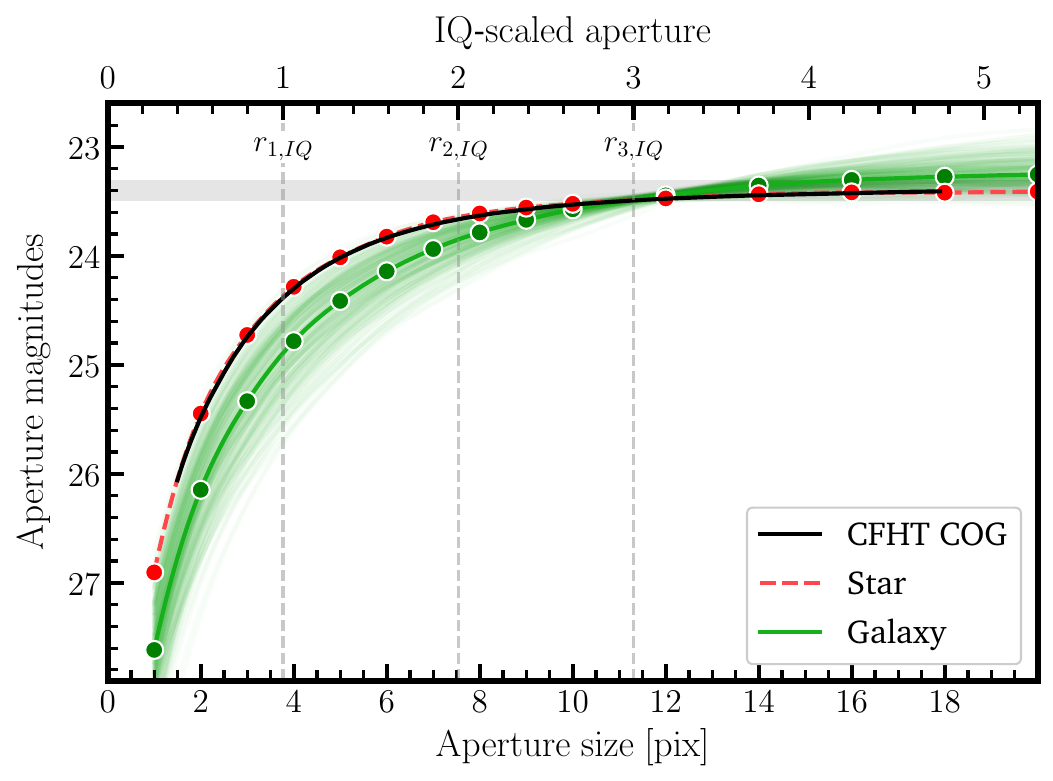}
\includegraphics[angle=0,clip,width=8.0cm]{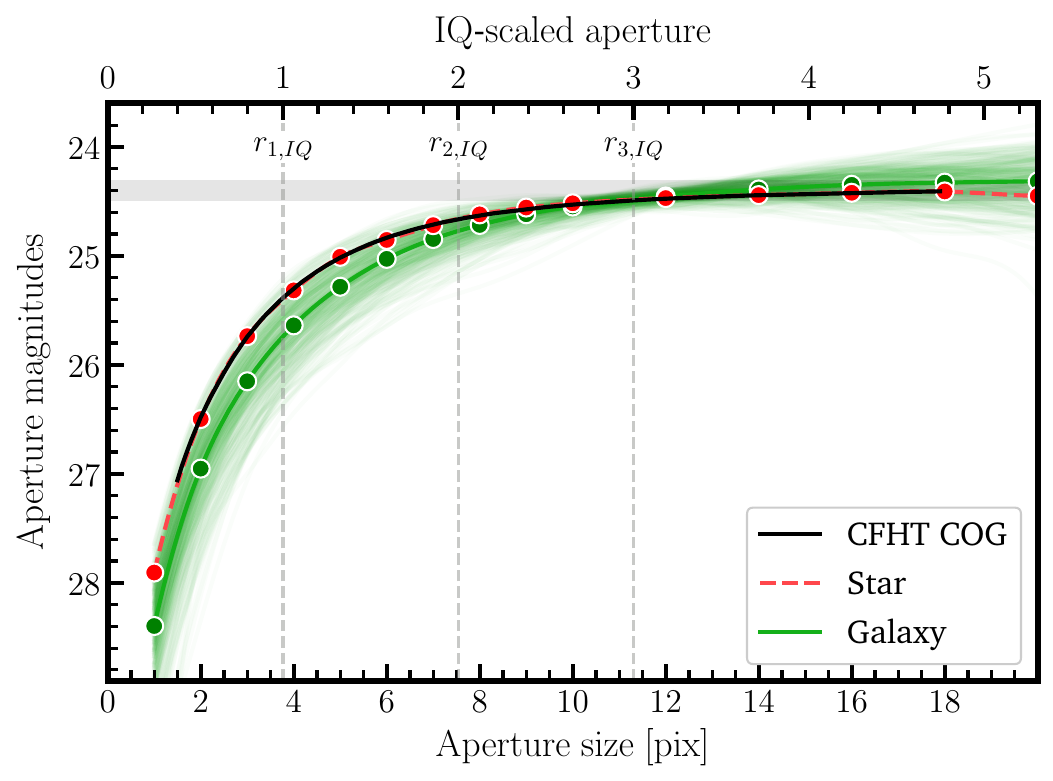}
\caption{Same as Fig.~\ref{fig:COG}, but for stars and galaxies around $r_\mathrm{COG} \sim 23.4$ (upper panel) and $r_\mathrm{COG} \sim 24.4$ (lower panel).}
\label{fig:append_SG}
\end{figure}

Fig.~\ref{fig:append_SG} is similar to Fig.~\ref{fig:COG}, but shows objects with $r_\mathrm{COG} \sim 23.4$ (upper panel) and $r_\mathrm{COG} \sim 24.4$ (lower panel). More specifically, the objects are selected with $0.69 < \mathrm{IQ} < 0.71$ and with COG magnitudes within 0.01 mag of $r_\mathrm{COG}=23.40$ and $r_\mathrm{COG}=24.40$. These selections correspond to 34 (231) and 36 (674) point sources (extended objects), respectively. For the point sources, the magnitude at each aperture varies very little, with a typical $1\sigma$ scatter of $\sim0.05$ mag. We therefore show only the median COG for the point sources, whereas for the extended objects, each line represents an individual galaxy, illustrating the much larger diversity in their COGs.

\section{Input parameter importance of the neural network}
\label{app:feat_imp}

\begin{figure*}
\includegraphics[angle=0,clip,width=16.5cm]{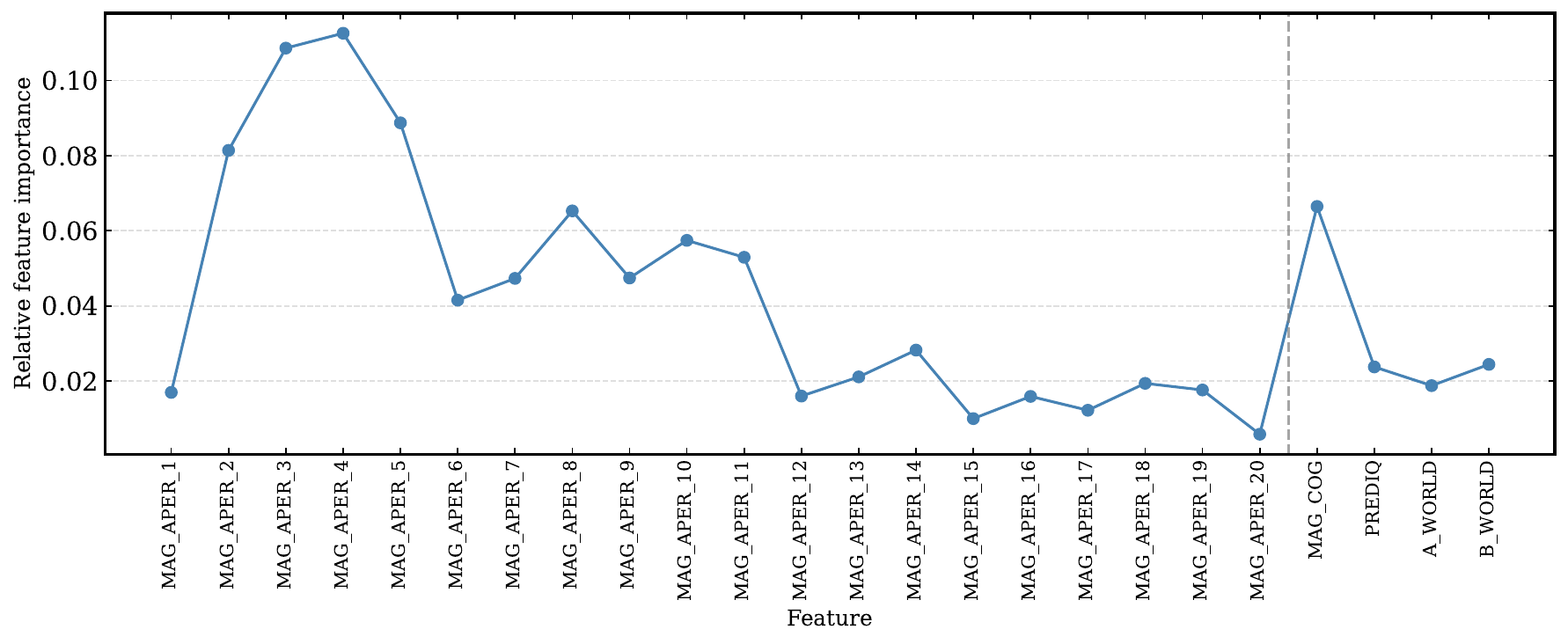}
\caption{Relative importance of each input parameter of the neural network method.}
\label{fig:imp_feat}
\end{figure*}

To quantify the relative contribution of each input feature to the classifier's decision, we computed SHAP (SHapley Additive exPlanations) values \citep{lundberg_2017}, which attribute to each input a value representing its marginal contribution to the model output, based on cooperative game theory. For each feature, we averaged the absolute SHAP value over all test samples and output classes, and normalised the result so that the sum over all features equals unity, yielding a relative importance fraction. Figure~\ref{fig:imp_feat} shows this quantity for each input feature. We find that aperture magnitudes measured beyond $\sim$12 pixels (corresponding to $2.2\,\mathrm{arcsec}$), together with the semi-major and semi-minor axis parameters, play a minor role, with a relative importance of less than 0.03, and contribute negligibly to the overall classification performance. This is expected, as aperture photometry at these larger radii is dominated by sky background and carries little additional discriminating information beyond what is already captured by smaller apertures. Nevertheless, since these quantities are readily available for the entire catalogue at negligible computational cost, we chose to retain them as inputs to the network.

\end{appendix}

\end{document}